# Anisotropic Raman scattering and lattice orientation identification of 2M-WS$_2$

Sabin Gautam[1,2,#], Sougata Mardanya[3,#], Joseph McBride[4], A K M Manjur Hossain[4], Qian Yang[5,2], Wenyong Wang[1,2], John Ackerman[6], Brian M. Leonard[4,2], Sugata Chowdhury[3,*] and Jifa Tian[1,2,*]

1. Department of Physics and Astronomy, University of Wyoming, Laramie, Wyoming, 82071, USA

2. Center for Quantum Information Science and Engineering, University of Wyoming, 82071, USA

3. Department of Physics, Howard University, Washington, DC, 20059, USA

4. Department of Chemistry, University of Wyoming, Laramie, Wyoming, 82071, USA

5. Center for Advanced Scientific Instrumentation, University of Wyoming, 82071, USA

6. Department of Chemical and Biological Engineering, University of Wyoming, Laramie, Wyoming, 82071, USA

*Email: sugata.chowdhury@howard.edu and jtian@uwyo.edu

ABSTRACT: Anisotropic materials with low symmetries hold significant promise for next-generation electronic and quantum devices. 2M-WS$_2$, a candidate for topological superconductivity, has garnered considerable interest. However, a comprehensive understanding of how its anisotropic features contribute to unconventional superconductivity, along with a simple, reliable method to identify its crystal orientation, remains elusive. Here, we combine theoretical and experimental approaches to investigate angle- and polarization-dependent anisotropic Raman modes of 2M-WS$_2$. Through first-principles calculations, we predict and analyze phonon dispersion and lattice vibrations of all Raman modes in 2M-WS$_2$. We establish a direct correlation between their anisotropic Raman spectra and high-resolution transmission electron microscopy images. Finally, we demonstrate that anisotropic Raman spectroscopy can accurately determine the crystal orientation and twist angle between two stacked 2M-WS$_2$ layers. Our findings provide insights into the electron-phonon coupling and anisotropic properties of 2M-WS$_2$, paving the way for the use of anisotropic materials in advanced electronic and quantum devices.

Two-dimensional (2D) transition metal dichalcogenides (TMDs) have emerged as a versatile platform for exploring exotic phenomena due to their diverse range of crystal structures and tunable electronic properties.[1,2] Among these, the *1T'* and $T_d$ phases, characterized by lattice distortions with lower crystal symmetry, exhibit pronounced anisotropic behavior.[3–7] This broken symmetry introduces novel physical properties,[8,9] including anisotropic charge transport[10] and enhanced spin-orbit coupling.[11,12] Thus, anisotropic TMDs with low symmetry in crystal structure represent a promising platform for investigating novel phenomena such as charge density waves[13] and topological phases.[14,15] Recently, 2M phase $WS_2$ (2M-$WS_2$),[16–24] characterized by its monoclinic crystal structure, exhibits remarkable anisotropic properties. For instance, vortex structures in 2M-$WS_2$ are highly elongated along *W-W* zigzag chains, with Majorana bound states exhibiting significant directional anisotropy.[17] Electrical transport reveals a pronounced crystal orientation-dependent conductivity and Seebeck coefficient.[24] Magnetotransport further highlights a strong anisotropic superconducting gap depending on the magnetic field direction.[23] Additionally, in-plane anisotropic plasmons have been detected in the far- and mid-infrared regimes.[18] Despite these advances, the anisotropic properties of 2M-$WS_2$, particularly in electron-phonon coupling and lattice vibration modes, being critical factors[22] in understanding its unconventional superconductivity, remain underexplored. In particular, anisotropic Raman scattering, a potential key to understanding the underlying physics of these phenomena, has yet to be comprehensively investigated.

Raman spectroscopy is a powerful tool for characterizing 2D materials, enabling the determination of layer thickness,[25] crystallographic orientations,[26] strain,[27] doping,[28] alloy composition,[29] and structural phase transition.[30,31] However, since the discovery of 2M-$WS_2$, there have been very limited studies investigating its Raman modes,[32,33] and a comprehensive analysis of its anisotropic properties remains lacking. Moreover, while bulk 2M-$WS_2$ preserves the inversion symmetry[34], breaking this symmetry in its bilayer can induce a Berry curvature dipole and a non-linear response,[35] suggesting that exotic quantum states may emerge in engineered structures of 2M-$WS_2$ with symmetry-breaking. For instance, the twist angle can modulate Josephson junction coupling between superconducting layers, offering a pathway to new quantum interference states arising from the moiré patterns formed in atomically thin layers. Therefore, developing a reliable method to determine the twist angle in anisotropic 2M-WS$_2$ is essential for unlocking its full potential in quantum device applications.

Here, we combine experimental and theoretical approaches to systematically investigate the anisotropic Raman modes of 2M-WS$_2$ thin layers using 532 nm and 638 nm excitation lasers. Employing first-principles calculations, we identify all observed Raman peaks of 2M-WS$_2$ corresponding to six $A_g$ and three $B_g$ modes. Angle-dependent polarized Raman spectroscopy reveals distinct symmetries in these modes, demonstrating the material's pronounced anisotropy. Furthermore, by coupling these findings with high-resolution transmission electron microscopy (HRTEM), we directly correlate the angle-dependent Raman spectra with the crystal orientation. This combination of experimental and theoretical methods allows us to demonstrate that polarized Raman spectroscopy can reliably determine the twist angle of 2M-WS$_2$/2M-WS$_2$ junctions.

The monoclinic structure (Figure 1a) of 2M-WS$_2$ belongs to the C$_{2/m}$ space group. Due to Peierls distortion,[16,31] the *W* atom deviates from the octahedral center and leads the *W-W* zigzag chain along the *b* direction. The zigzag chains facilitate two distances, 0.23 nm and 0.34 nm, along the *c*-direction. 2M-WS$_2$ crystals (Figure S1a) were synthesized through a synthetic method that we developed previously.[30,36] Figure 1b shows the optical image of an exfoliated few-layer 2M-WS$_2$ with varying thickness on a SiO$_2$/Si substrate, showing a clear thickness-dependent optical contrast. The crystal structure of bulk 2M-WS$_2$ was confirmed by X-ray diffraction (XRD) (Figure 1c) with a = 12.85 Å, b = 3.22 Å, c = 5.70 Å, α = γ = 90°, and β = 112.91°. To determine its chemical composition, we carried out the energy dispersive X-ray (EDX) mapping (Figure S1) and determined the atomic ratio of *W*: *S* (34.12%: 65.72%) is very close to 1:2.

To experimentally characterize and theoretically understand the lattice vibration modes in 2M-WS$_2$, we employed Raman spectroscopy and first-principle calculations. Figure 1d shows the schematic of our Raman spectroscope for studying the polarization and angle-dependences of Raman modes of 2M-WS$_2$. Figure 1e shows the characteristic Raman spectra of a few-layer 2M-WS$_2$ (Figure 1b) using circularly polarized 532 and 638 nm lasers. With the 532nm laser, we identified seven distinct Raman peaks spanning from 80 to 500 cm$^{-1}$. These include two $B_g$ modes at 111 and 239 and five $A_g$ modes at 126, 178, 268, 317, and 407 cm$^{-1}$. In contrast, Raman spectra acquired with the 638 nm laser only reveal six peaks and the Raman peak at 407 cm$^{-1}$ exhibits a marked reduction or completely disappears. We attribute this observation to the diminished electron-photon resonance when the excitation energy shifts from 532 nm to 638 nm. Conversely, the Raman peak at 126 cm$^{-1}$ is significantly enhanced under the 638 nm laser. These findings underscore a pronounced laser wavelength dependence of Raman modes in few-layer 2M-WS$_2$ based on the optical transition selection rule (Supporting Information).

We employed density functional perturbation theory to predict and gain a comprehensive understanding of the Raman active modes for 2M-WS$_2$. The phonon band structure for the conventional cell is presented in Figure 2a. At the high symmetry point Γ, phonon bands can be characterized by their irreducible representations (IRRs) of the point group symmetry $C_{2h}$ (2/m). Among these, the vibrational modes with IRRs $A_g$ and $B_g$ transform as the respective polarizability tensor component ($x^2, y^2, z^2, xz$) and ($xy, yz$) are identified as the Raman active modes (Table S1). We calculated the electron-phonon coupling resolved phonon band structure as presented in Figure S2. These results indicate an enhancement of electron-phonon coupling for the Raman active modes at the high-symmetry $\Gamma$ point. For 2M-WS$_2$, the $A_g$ and $B_g$ modes are highlighted in the phonon spectrum of Figure 2a by the corresponding *ω* values in green and red colors, respectively. We further calculated the polarization-dependent intensities of resonance Raman spectra using the dipole vector and electron-phonon matrix elements from the density functional theory (DFT) calculations implemented in the QERaman package, as shown in Figure 2b. The solid and dotted lines represent the intensities of the Raman modes with the incident laser at 638 nm and 532 nm, respectively. It is evident that $B_g$ modes have appeared only for the XY polarization (red line), while $A_g$ modes have finite peaks for XX and ZZ polarization (green and blue lines). We further plotted the atomic displacements of the lattice vibrations corresponding to each Raman mode in Figures 2c-d. The vibrations for $B_g$ modes (Figure 2c) appear to be predominantly along the Y-direction (along the in-plane crystal *b* axis), causing a change in polarizability along the X-direction. Meanwhile, the atom displacements for $A_g$ modes in Figure 2d are mostly restricted to the XZ plane (in the *a-c* plane) with obvious out-of-plane components.

To explore the in-plane anisotropy of vibration modes, we conducted polarization-dependent Raman spectroscopy under both parallel and perpendicular configurations. Figure 1d shows the schematic diagram of the Raman setup. The incident light is consistently polarized along the horizontal direction (Y) using a polarizer, and a Raman analyzer placed before the spectrometer enables analysis of the scattered light, either parallel or perpendicular to the polarization of the incident light. Angle-dependent polarized Raman signals were collected with the sample rotated counterclockwise in the X-Y plane (Figure 1d). When the 2M-WS$_2$ flake has six or more layers, the frequencies of $A_g$ and $B_g$ Raman modes converge to the bulk values.[33] The intensity of Raman peaks depends on the excitation laser wavelengths and the crystal orientation with respect to the laser polarization, showing strong anisotropy. Here, we take an 8.5 nm thick sample (Figure 3a) to study the anisotropic Raman response and to further establish its relationship with crystal orientation. We define the Y-axis as the 0-degree reference (Figure 3a) which is parallel to the flake's *b*-axis, being the same direction as the polarization vector of the incident laser. In parallel configuration, the excitation laser is parallel to the Y-axis, and the horizontal Raman polarization allows the scattered signal to pass parallel to the sample. The polarized Raman scattering intensities evolving with the sample rotation from 0° to 360° under parallel and perpendicular configurations are shown in Figures 3b and c, respectively. The peak intensities of Raman modes of the 2M-WS$_2$ layer exhibit distinct periodicities (Figures 3 and S3), indicating pronounced in-plane anisotropy. Figures 3d,e show the polar plots of the Raman peak intensities of both $B_g$ and $A_g$ modes as a function of the rotation angle. In both configurations, we see that $B_g$ modes located at 111 and 239 cm$^{-1}$ exhibit a four-fold symmetry. We also observe that the Raman modes can show distinct anisotropy under the same laser excitation. For instance, under 532 nm laser excitation in the parallel configuration (Figures 3b and e), $A_g$ modes at 126 and 407 cm$^{-1}$ reach maximum intensities at 0°, while the mode at 268 cm$^{-1}$ shows maximum intensity at 90°. The two $B_g$ modes exhibit similar angle-dependent Raman intensity with four-fold symmetry, showing maximum intensities at 45° and minimum at 0° and 90°, while the three $A_g$ modes (at 126, 268 and 407 cm$^{-1}$) display two-fold rotational symmetry and the mode at 317 cm$^{-1}$ exhibits four-fold symmetry. In the perpendicular configuration (Figures 3c and S3b), we observe six Raman active modes with four $A_g$ modes (126, 268, 317 and 407 cm$^{-1}$) and two $B_g$ modes (111 and 239 cm$^{-1}$). Furthermore, the Raman mode at 178 cm$^{-1}$ is absent in this configuration but retains nearly constant intensity in the parallel configuration (Figures 3b, c and S4a), consistent with its Raman tensor having symmetric diagonal elements and near zero off-diagonal components. The two $B_g$ modes exhibit similar angle-dependent Raman intensity with four-fold symmetry. In contrast, three $A_g$ modes (at 126, 268, and 317 cm$^{-1}$) display the four-fold rotational symmetry, while the symmetry of the mode at 407 cm$^{-1}$ is two-fold.

To further explore the effect of laser wavelengths, we investigated the angle dependence of the polarized Raman spectra using the 638 nm laser. Specifically, in the parallel configuration (Figures 4a, c, d, S4a and 5b), Raman modes with $A_g$ symmetry at 126, 178, 268 and 317 cm$^{-1}$ exhibit maximum intensities at ~90°, while Raman mode at 407 cm$^{-1}$ shows a maximum intensity around 0°. When comparing results to those obtained with the 532 nm laser in the same configuration, we observed that the anisotropy of certain Raman modes changes as the laser wavelength varies, with some modes becoming isotropic under specific laser excitation. For instance, the $A_g$ mode at 178 cm$^{-1}$, showing pronounced anisotropy under the 638 nm laser (Figures 4a, S4a and 5b), becomes nearly isotropic with the 532 nm laser excitation (Figures 3b and S5a). This phenomenon can be attributed to the optical transition selection rule[37] (see detailed discussion in the supporting information). Notably, using the 638 nm laser in a perpendicular configuration (Figures 4b-d, S4b and 5b), we observe all nine theoretically predicted Raman active modes. Among these, 111, 239, and 276 cm$^{-1}$ correspond to $B_g$ modes, while the others are $A_g$ modes. Although the Raman peaks at 276 and 333 cm$^{-1}$ are too weak to be included in the polar plots, they are evident in the Raman intensities color plot (Figure 4b) and the line plots (Figure S4b).

To understand the polarization dependence of the Raman spectra, we can use the classical theory of polarized Raman spectrum to obtain the following intensity equation,

$$I_R \propto |\langle e_i|R_{IRR}|e_s\rangle|^2.$$

Here $|e_i\rangle$ and $|e_s\rangle$ are the polarization vectors of the incident, and scattered light and $R_{IRR}$ is the Raman tensor of the corresponding modes identified by the IRRs. In our case, considering the geometry of the 2M-WS$_2$ unit cell, the Raman tensors for the Raman active modes ($A_g$ and $B_g$) in the complex forms are

$$R_{A_g} = \begin{pmatrix} ae^{i\phi_a} & 0 & ce^{i\phi_c} \\ 0 & be^{i\phi_b} & 0 \\ ce^{i\phi_c} & 0 & de^{i\phi_d} \end{pmatrix}, \quad R_{B_g} = \begin{pmatrix} 0 & ee^{i\phi_e} & 0 \\ ee^{i\phi_e} & 0 & fe^{i\phi_f} \\ 0 & fe^{i\phi_f} & 0 \end{pmatrix},$$

where $a, b, c, d, e$ and $f$ represent the amplitude of Raman tensor elements, respectively. For the backscattering geometry with parallel polarization, the polarization vectors are given by $\langle e_i| = \langle e_s| = (\cos\theta \quad \sin\theta \quad 0)$. The Raman intensity is then given by,

$$I_{A_g}^{\parallel} = a^2 \cos^4\theta + b^2 \sin^4\theta + 2ab\cos^2\theta \sin^2\theta \cos\phi_{ab},$$

$$I_{B_g}^{\parallel} = 4e^2 \cos^2\theta \sin^2\theta,$$

where $\phi_{ab} = (\phi_a - \phi_b)$, is the relative phase between $a$ and $b$. For the perpendicular polarization, the polarization vectors for incident and scattered light are $\langle e_i| = (\cos\theta \quad \sin\theta \quad 0)$, and $\langle e_s| = (-\sin(\theta) \cos(\theta) 0)$. With this assumption, the intensity equations for perpendicular polarization can be formulated as,

$$I_{A_g}^{\perp} = a^2 \cos^2\theta \sin^2(\theta) + b^2 \cos^2(\theta) \sin^2\theta - 2ab\cos\theta\sin\theta \cos(\theta) \sin(\theta) \cos\phi_{ab},$$

$$I_{B_g}^{\perp} = e^2(\cos^2\theta \cos^2(\theta) + \sin^2\theta \sin^2(\theta) - 2\cos\theta\sin\theta \cos(\theta) \sin(\theta) \cos\phi_{ab}).$$

We noticed that for some $A_g$ modes the experimental data can fit significantly better when an additional angle $\delta$ is introduced into the polarization vector as $\langle e_s| = (-\sin(\theta + \delta) \cos(\theta + \delta) 0)$. The equation can be modified as follows,

$$I_{A_g}^{\perp} = a^2 \cos^2\theta \sin^2(\theta + \delta) + b^2 \cos^2(\theta + \delta) \sin^2\theta - 2ab\cos\theta\sin\theta \cos(\theta + \delta) \sin(\theta + \delta) \cos\phi_{ab},$$

We use these equations to fit the experimental data as depicted in Figures 3d-e and 4c. We note that $\delta$ is introduced to account for the existing misalignment in the Raman system. The parameters obtained from the fittings are listed in supplementary Tables S2 and 3.

To determine whether anisotropic Raman scattering is influenced by layer thickness, we carried out angle-dependent polarized Raman spectroscopy measurements on ultra-thin flakes (~2 and 4.2 nm) and compared them with 12 nm flake under both the parallel and perpendicular configurations (Figures S7-9, Tables S4 and S5) using the 532 nm laser. We observed a clear thickness dependence of the Raman modes (Figure S8c), consistent with a previous study on 2M-WS$_2$ samples thinner than ~ 5.5 nm[33]. Notably, in these ultra-thin samples, the first $B_g$ mode splits into two with similar symmetry (Tables S4 and 5), possibly due to the Davydov splitting, observed in other materials like WTe$_2$.[38] We further note that the angle-dependent polarized Raman spectra of 2 nm flake exhibit a similar angular response to those of the 4.2, 8.5, and 12 nm flakes. These findings provide insights into the understanding of anisotropic Raman properties of 2M-WS$_2$ and pave the pathway to determine its crystal orientation using Raman spectroscopy.

To correlate the anisotropic Raman spectroscopy of 2M-WS$_2$ with its lattice structure, we conducted HRTEM, selected area electron diffraction (SAED), and polarized Raman spectroscopy on the same sample. Figure 5a shows the TEM image of the 2M-WS$_2$ layer on a copper (Cu) TEM grid. The SAED pattern of the 2M-WS$_2$ flake (Figure 5b) further demonstrates the 2M phase. Figure 5c is an HRTEM image taken from the orange box in Figure 5a, revealing the *W-W* zigzag chains along the *b*-direction. We further identified the non-uniform distances between *W* atom lines, $d_1$ = 2.36 Å and $d_2$ = 3.37 Å, consistent with the previously reported results of 2M-WS$_2$.[33] The corresponding Fast Fourier Transform (FFT) of the HRTEM image shows the characteristic rectangular pattern of the 2M phase. To confirm the single-crystalline nature of the flake, we obtained HRTEM images from the top-right edge of the flake (Figure S10). A comparison with images from the bottom edge reveals consistent crystal structure across the entire flake, as both HRTEM images and their corresponding FFTs displayed the same *W* atom configurations. Furthermore, the SAED patterns closely match the XRD results of 2M-WS$_2$, conclusively confirming its crystal orientation and the *b*-axis in the flake.

We further collected the angle-dependent polarized Raman spectra of this flake in a parallel configuration using the 532 nm laser. The Raman false-color plot (Figure S11) exhibits the same angle dependence as those observed in the 8.5 nm-thick 2M-WS$_2$ flake (Figure 3b). For instance, the maximum intensity of the $A_g$ mode at 268 cm$^{-1}$ with two-fold symmetry (Figure 5d) occurs when the excitation laser is polarized perpendicular to the *W-W* chain direction using both 532 and 638 nm lasers. We note that the Raman signal of the samples on the Cu grid is generally weaker. These results demonstrate that the *b* direction can be accurately determined by identifying the maximum Raman intensity through angle-dependent Raman spectroscopy measurements. Thus, by directly correlating the HRTEM and angle-dependent Raman spectra of the same flake, we established a simple, reliable method for determining the crystal orientation of 2M-WS$_2$ thin layers.

We employed angle-dependent polarized Raman spectroscopy to determine the twist angle of a junction made of two twisted 2M-WS$_2$ layers. To prepare the junction, we first exfoliated a 2M-WS$_2$ thin flake onto a SiO$_2$/Si substrate and then transferred another flake to the top using the dry transfer technique. Figure 5e shows the optical image of the junction with a twist angle, acting as a Josephson junction. We performed angle-dependent polarized Raman measurements on two layers of the junction in a parallel configuration using the 638 nm laser. Figures S12a-c show the color plots of the Raman spectra of the top flake, bottom flake, and junction, respectively. The $A_g$ Raman modes at 178 cm$^{-1}$ shows a two-folded periodicity, with the maximum intensity aligned along the *b*-axis of the corresponding layer. Figure 5f shows the polar plots of the $A_g$ mode at 178 cm$^{-1}$ collected from the bottom flake (black), top flake (red) and the overlapped region (grey). By comparing the angles at which the Raman mode reaches their maximum intensity, we can determine the twist angle to be approximately 10°. We note that the same result can be obtained by analyzing the Raman mode at 268 cm$^{-1}$ (Figure S12d). Notably, the polarized Raman intensities within the junction region (Figures 5f and S12c) are largely enhanced, primarily due to variations in the thickness of 2M-WS$_2$. We speculate that as the thickness of the two layers approaches the monolayer or bilayer regime, more distinct Raman features emerge in the junction region. Our finding paves the way for further exploration of moiré patterns and novel quantum states in twisted 2M-WS$_2$ flakes as well as Josephson couplings using polarized Raman spectroscopy.

In summary, we present the first angle-dependent Raman spectroscopy measurements on 2M-WS$_2$, complemented by the theoretical identification of all nine Raman active modes using DFT calculations. By integrating HRTEM with anisotropic Raman spectroscopy, we accurately determine the edge orientation of 2M-WS$_2$. Our methodology enables the precise determination of twist angles in 2M-WS$_2$/2M-WS$_2$ junctions. This study provides insights into the electron-phonon coupling and the understanding of anisotropic Raman vibrational modes in 2M-WS$_2$ and demonstrates the potential of angle-dependent Raman spectroscopy for exploring moiré patterns and Josephson coupling in twisted 2M-WS$_2$ structures and Junctions.

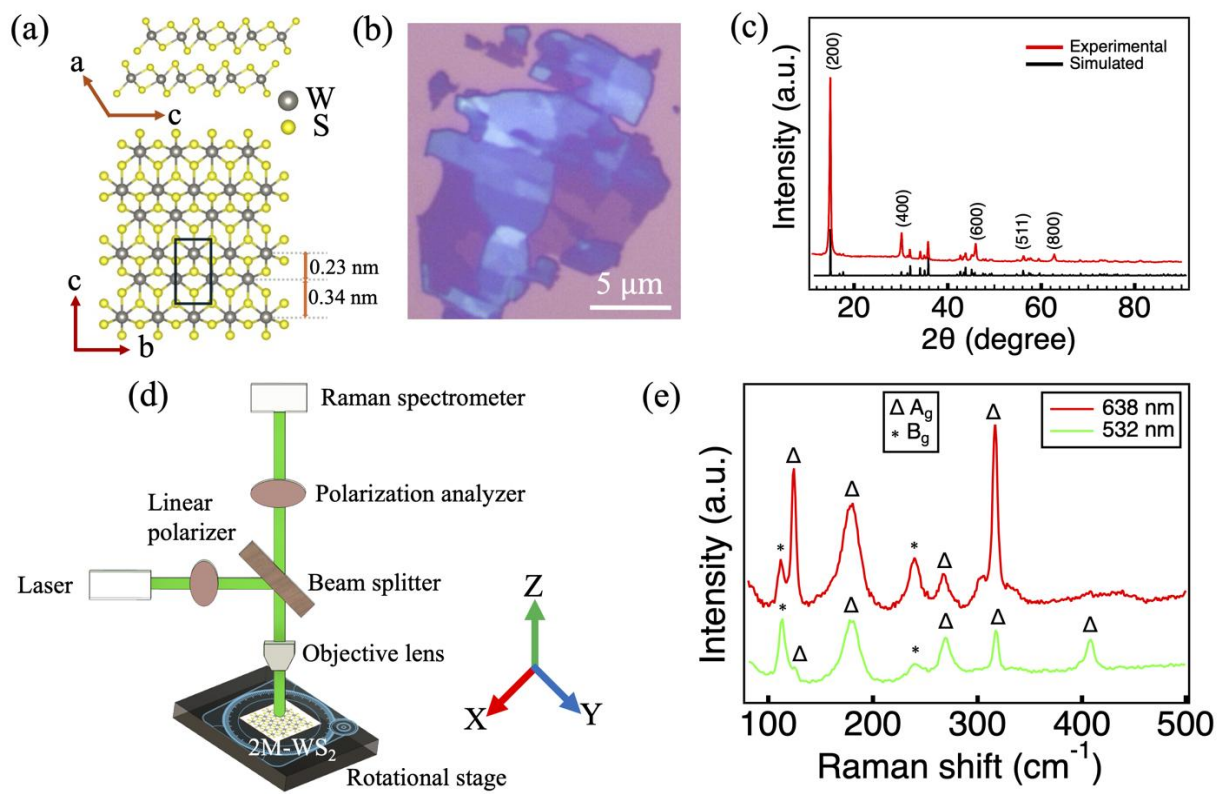

**Figure 1.** Characterization of bulk and few-layer 2M-WS$_2$. (a) Schematic of the crystal structure of 2M-WS$_2$. Top panel: side view; bottom panel: top view. The black rectangle outlines the unit cell. (b) Optical image of a 2M-WS$_2$ flake with varying thickness on a SiO$_2$/Si substrate. (c) XRD pattern of 2M-WS$_2$ crystals, where the black lines were simulated from its single crystal structure. (d) Schematic of the angle-dependent polarized Raman spectroscopy. The incident laser is linearly polarized. (e) Raman spectra of a few-layer 2M-WS$_2$ using 532 and 638 nm laser excitations. The $A_g$ and $B_g$ modes are shown with the symbols ∆ and ∗, respectively.

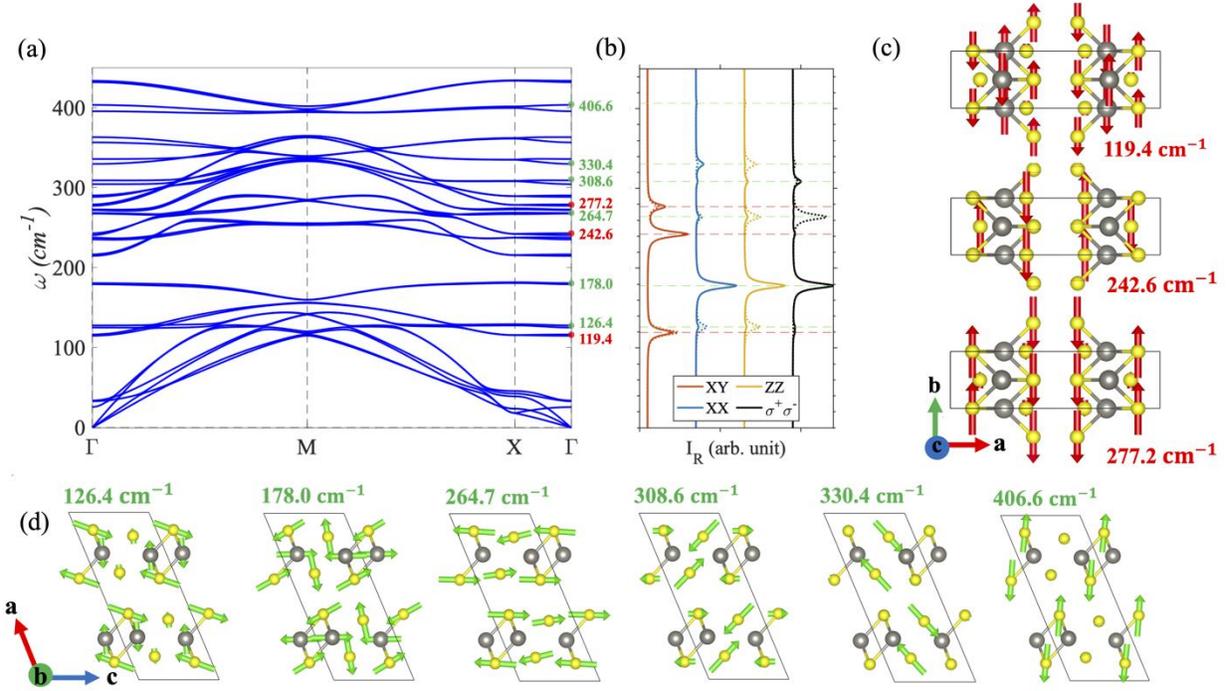

**Figure 2.** Phonon band structure and lattice vibrations in 2M-WS$_2$. (a) Phonon dispersion of 2M-WS$_2$ along the high symmetry direction in the Brillouin zone. The $A_g$ and $B_g$ Raman active modes at $\Gamma$ point are heightened by the green and red color, respectively by the side of the spectrum. (b) The intensity variation of different Raman modes with different polarization is represented by different colors, while the dashed and continuous lines correspond to 532 nm and 638 nm laser wavelengths, respectively. (c) Visual representation of the real space lattice vibrations corresponding to the three $B_g$ modes showing intra-layer components aligned purely along the b-direction. (d) Lattice vibrations corresponding to the six $A_g$ modes are showing out of layer components.

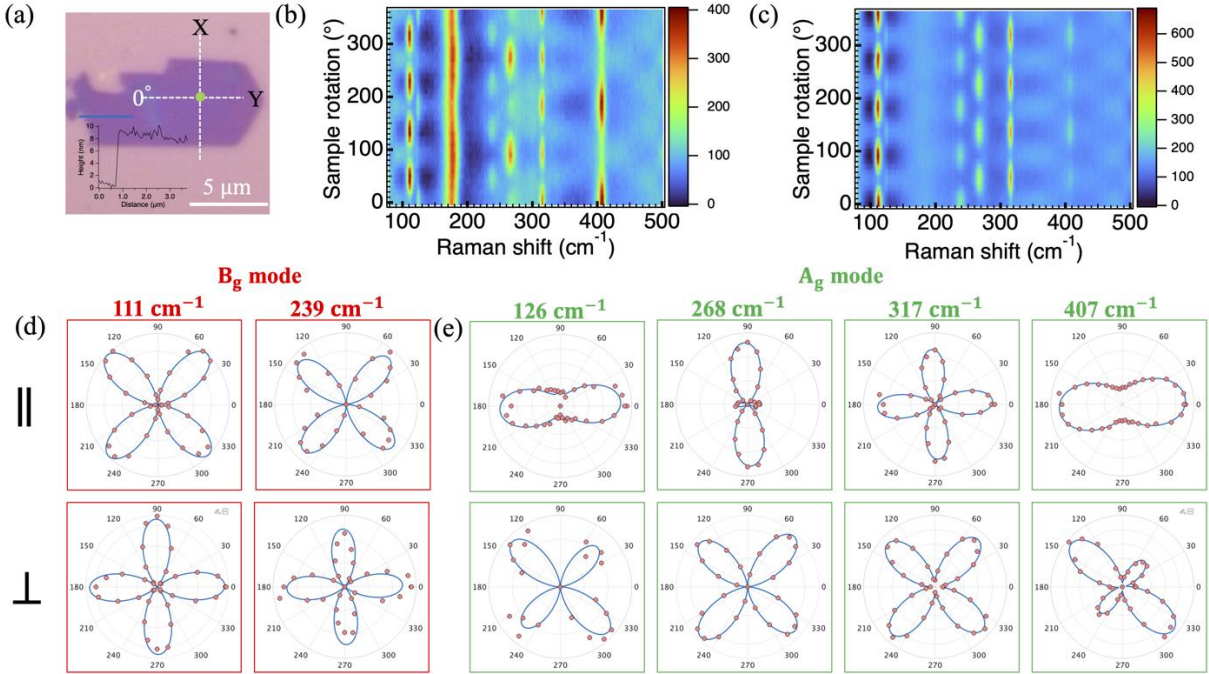

**Figure 3.** Angle dependence of polarized Raman response under parallel (∥) and perpendicular (⊥) configurations using a 532 nm excitation laser. (a) Optical image of a few-layer (8.5 nm) 2M-$WS_2$ flake on a $SiO_2$/Si substrate. The flake is rotated counterclockwise while taking the angle-dependent Raman spectra. The green dot indicates where the Raman measurement was performed. Inset is the line (black) profile of the flake labeled with a blue line acquired by AFM. False-color plots of polarized Raman intensities under parallel (b) and perpendicular (c) configurations. Polar plots of the peak intensities of $B_g$ modes (d) and $A_g$ modes (e) as a function of sample rotation angle under ∥ and ⊥ configurations, respectively. The b-axis is defined as the Y- direction in the experimental coordinate, the X- direction is perpendicular to the *b*-axis, and θ is the angle between the directions of the *b*-axis and the incident polarization.

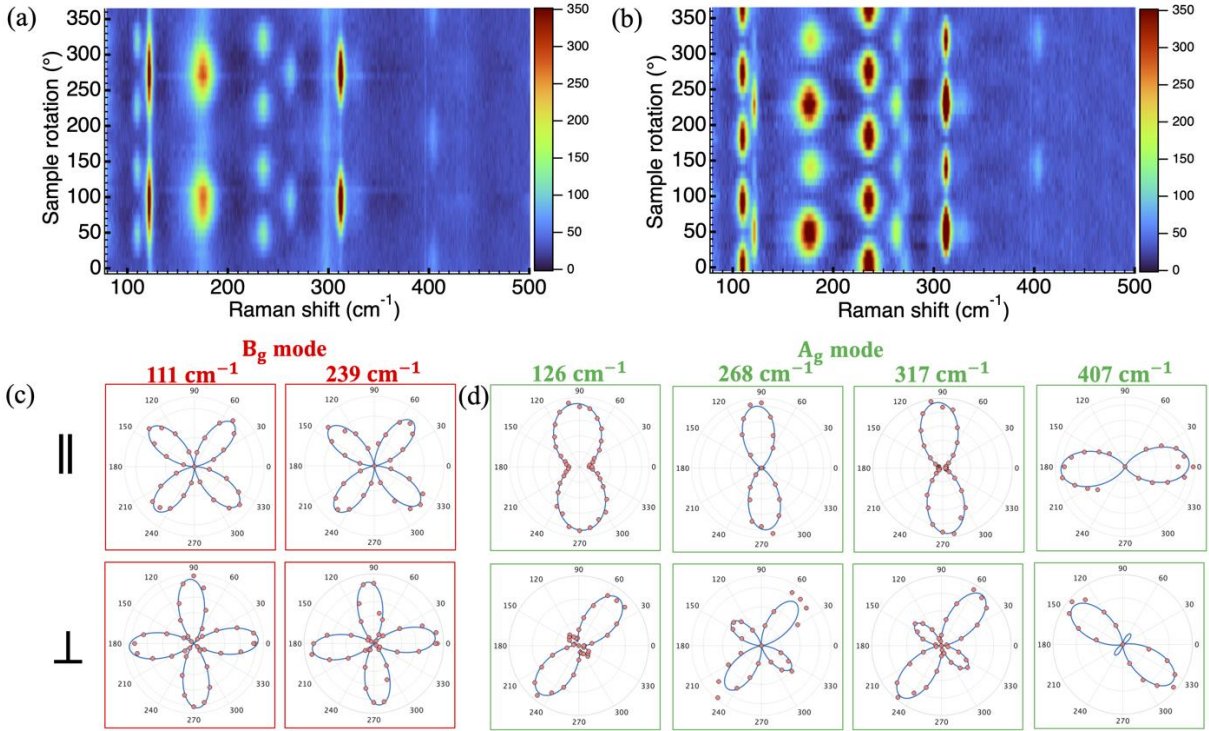

**Figure 4.** Angle dependence of polarized Raman response under parallel (∥) and perpendicular (⊥) configuration using a 638 nm excitation laser. The false-color plots of polarized Raman spectra under parallel (a) and perpendicular (b) configurations. Polar plots of peak intensities of $B_g$ modes (c) and $A_g$ modes (d) as functions of sample rotation angle under ∥ and ⊥ configurations, respectively.

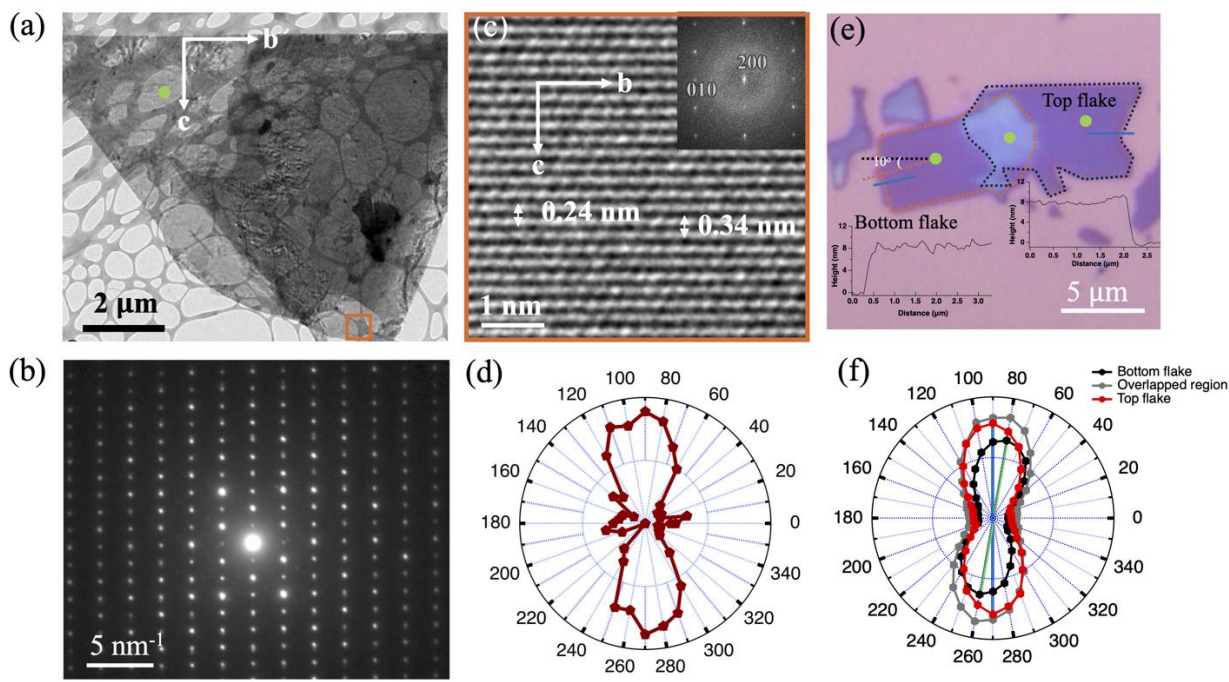

**Figure 5.** Correlation between TEM and Raman spectroscopy and determination of twist angle of a 2M-$WS_2$/2M-$WS_2$ junction. (a) Low magnification TEM image of a few-layer 2M-$WS_2$ on a Cu grid. (b) Selected area electron diffraction (SAED) pattern of the flake. (c) High-resolution TEM (HRTEM) image and corresponding fast Fourier transform (FFT) pattern of the image taken from the orange square area in (a). (d) Polar plot of the $A_g$ mode at 268 cm$^{-1}$ of the same 2M-$WS_2$ flake in (a) under parallel configuration using the 532 nm laser. (e) Optical image of a 2M-$WS_2$/2M-$WS_2$ junction. Insets are the line (black) profiles of the flakes labeled with blue lines acquired by AFM. The green dots mark the locations where the Raman measurements were performed. (f) Polar plots of the $A_g$ mode at 178 cm$^{-1}$ for the bottom flake (black), top flake (red), and overlapped region (grey). The blue and green lines indicate the direction of maximum intensities for the top and bottom flake, respectively. The twist angle is approximately 10°.

**Supporting Information**. The Supporting Information is available free of charge at

Methods and figures include SEM, EDX mapping of 2M-W$_2$ crystals, angular dependence of polarized Raman spectra of 2, 4.2, 8.5 and 12 nm flakes under parallel and perpendicular configurations using different laser excitations, analysis of laser energy dependence and optical transition selection rule, electron-phonon coupling-weighted phonon band structure, theoretical calculations for phonon modes in 2M-WS$_2$ at $\Gamma$ point, Fitting parameters for both $A_g$ and $B_g$ modes.

**Author Contributions**

**Acknowledgment**

This research was mainly supported by the U.S. Department of Energy, Office of Basic Energy Sciences, Division of Materials Sciences and Engineering under award DE-SC0021281 for sample and device fabrication and after its completion by DE-SC0024188 for Raman measurements. J.T. also acknowledges the financial support of the U.S. National Science Foundation (NSF) grant 2228841 and 2327410 for data analysis. This work was also supported by the Science Institute's PhD Fellows Program at the University of Wyoming. S. M. and S. C. from Howard University, work supported by the U.S. Department of Energy, Office of Science, Basic Energy Sciences Grant No. DE-SC0022216. This research at Howard University used the resources of Accelerate ACCESS PHYS220127 and PHYS2100073.

**References**

1. Manzeli, S., Ovchinnikov, D., Pasquier, D., Yazyev, O. V. & Kis, A. 2D transition metal dichalcogenides. *Nat. Rev. Mater.* **(2017)**, 2, 17033.

2. Wang, Q. H., Kalantar-Zadeh, K., Kis, A., Coleman, J. N. & Strano, M. S. Electronics and optoelectronics of two-dimensional transition metal dichalcogenides. *Nat. Nanotechnol.* **(2012)**, 7, 699–712.

3. Sokolikova, M. S. & Mattevi, C. Direct synthesis of metastable phases of 2D transition metal dichalcogenides. *Chem. Soc. Rev.* **(2020)**, 49, 3952–3980.

4. Lai, Z. *et al.* Metastable 1T′-phase group VIB transition metal dichalcogenide crystals. *Nat. Mater.* **(2021)**, 20, 1113–1120.

5. Zhao, X. *et al.* Raman Spectroscopy Application in Anisotropic 2D Materials. *Adv. Electron. Mater.* **(2024)**, 10, 2300610.

6. Gong, C. *et al.* Electronic and Optoelectronic Applications Based on 2D Novel Anisotropic Transition Metal Dichalcogenides. *Adv. Sci.* **(2017)**, 4, 1700231.

7. Liu, L. *et al.* Phase-selective synthesis of 1T′ $MoS_2$ monolayers and heterophase bilayers. *Nat. Mater.* **(2018)**, 17, 1108–1114.

8. Tang, S. *et al.* Quantum spin Hall state in monolayer 1T'-$WTe_2$. *Nat. Phys.* **(2017)**, 13, 683–687.

9. Muechler, L., Alexandradinata, A., Neupert, T. & Car, R. Topological Nonsymmorphic Metals from Band Inversion. *Phys. Rev. X* **(2016)**, 6, 041069.

10. Nam, G. *et al.* In-Plane Anisotropic Properties of 1T′-$MoS_2$ Layers. *Adv. Mater.* **(2019)**, 31, 1807764.

11. Qian, X., Liu, J., Fu, L. & Li, J. Quantum spin Hall effect in two-dimensional transition metal dichalcogenides. *Science* **(2014)**, 346, 1344–1347.

12. Song, Y.-H. *et al.* Observation of Coulomb gap in the quantum spin Hall candidate single-layer 1T'-$WTe_2$. *Nat. Commun.* **(2018)**, 9, 4071.

13. Hovden, R. *et al.* Atomic lattice disorder in charge-density-wave phases of exfoliated dichalcogenides (1T-$TaS_2$). *Proc. Natl Acad. Sci.* **(2016)**, 113, 11420–11424.

14. Bruno, F. Y. *et al.* Observation of large topologically trivial Fermi arcs in the candidate type-II Weyl semimetal WTe$_2$. *Phys. Rev. B* **(2016)**, 94, 121112.

15. Pan, H., Xie, M., Wu, F. & Das Sarma, S. Topological Phases in AB-Stacked MoTe$_2$/WSe$_2$: Z$_2$ topological Insulators, Chern Insulators, and Topological Charge Density Waves. *Phys. Rev. Lett.* **(2022)**, 129, 056804.

16. Fang, Y. *et al.* Discovery of Superconductivity in 2M WS$_2$ with Possible Topological Surface States. *Adv. Mater.* **(2019)**, 31, 1901942.

17. Yuan, Y. *et al.* Evidence of anisotropic Majorana bound states in 2M-WS$_2$. *Nat. Phys.* **(2019)**, 15, 1046–1051.

18. Xing, Q. *et al.* Tunable anisotropic van der Waals films of 2M-WS$_2$ for plasmon canalization. *Nat. Commun.* **(2024)**, 15, 2623.

19. Li, Y. W. *et al.* Observation of topological superconductivity in a stoichiometric transition metal dichalcogenide 2M-WS$_2$. *Nat. Commun.* **(2021)**, 12, 2874.

20. Che, X. *et al.* Gate-Tunable Electrical Transport in Thin 2M-WS$_2$ Flakes. *Adv. Electron. Mater.* **(2019)**, 5, 1900462.

21. Wang, L. S. *et al.* Nodeless superconducting gap in the topological superconductor candidate 2M–WS$_2$ *Phys. Rev. B* **(2020)**, 102, 024523.

22. Lian, C.-S., Si, C. & Duan, W. Anisotropic Full-Gap Superconductivity in 2M-WS$_2$ Topological Metal with Intrinsic Proximity Effect. *Nano Lett.* **(2021)**, 21, 709–715.

23. Zhang, E. *et al.* Spin–orbit–parity coupled superconductivity in atomically thin 2M-WS$_2$. *Nat. Phys.* **(2023)**, 19, 106–113.

24. Yang, Y. *et al.* Anomalous enhancement of the Nernst effect at the crossover between a Fermi liquid and a strange metal. *Nat. Phys.* **(2023)**, 19, 379–385.

25. Zhang, X. *et al.* Phonon and Raman scattering of two-dimensional transition metal dichalcogenides from monolayer, multilayer to bulk material. *Chem. Soc. Rev.* **(2015)**, 44, 2757–2785.

26. Wang, Y., Cong, C., Qiu, C. & Yu, T. Raman Spectroscopy Study of Lattice Vibration and Crystallographic Orientation of Monolayer MoS$_2$ under Uniaxial Strain. *Small* **(2013)**, 9, 2857–2861.

27. Mohiuddin, T. M. G. *et al.* Uniaxial strain in graphene by Raman spectroscopy: G peak splitting, Grüneisen parameters, and sample orientation. *Phys. Rev. B* **(2009)**, 79, 205433.

28. Mueller, N. S. *et al.* Evaluating arbitrary strain configurations and doping in graphene with Raman spectroscopy. *2D Mater.* **(2017)**, 5, 015016.

29. Zhang, M. *et al.* Two-Dimensional Molybdenum Tungsten Diselenide Alloys: Photoluminescence, Raman Scattering, and Electrical Transport. *ACS Nano* **(2014)**, 8, 7130–7137.

30. Gautam, S. *et al.* Controllable superconducting to semiconducting phase transition in topological superconductor 2M-$WS_2$. *2D Mater.* **(2024)**, 11, 015018.

31. Duerloo, K.-A. N., Li, Y. & Reed, E. J. Structural phase transitions in two-dimensional Mo- and W-dichalcogenide monolayers. *Nat. Commun.* **(2014)**, 5, 4214.

32. Li, Y. *et al.* Evidence of strong and mode-selective electron–phonon coupling in the topological superconductor candidate 2M-$WS_2$. *Nat. Commun.* **(2024)**, 15, 6235.

33. Liu, X. *et al.* High intrinsic phase stability of ultrathin 2M $WS_2$. *Nat. Commun.* **(2024)**, 15, 1263.

34. Cho, S. *et al.* Direct Observation of the Topological Surface State in the Topological Superconductor 2M-$WS_2$. *Nano Lett.* **(2022)**, 22, 8827–8834.

35. Joseph, N. B. & Narayan, A. Topological properties of bulk and bilayer 2M $WS_2$: a first-principles study. *J. Phys.Condens. Matter* **(2021)**, 33, 465001.

36. Samarawickrama, P. *et al.* Two-Dimensional 2M-$WS_2$ Nanolayers for Superconductivity. *ACS Omega* **(2021)**, 6, 2966–2972.

37. Huang, S. *et al.* In-Plane Optical Anisotropy of Layered Gallium Telluride. *ACS Nano* **(2016)**, 10, 8964–8972.

38. Cao, Y. et al. Anomalous vibrational modes in few layer $WTe_2$ revealed by polarized Raman scattering and first-principles calculations. *2D Mater.* **(2017)**, 4, 035024.